\UseRawInputEncoding
\documentclass[journal=jacsat,manuscript=article]{achemso}

\usepackage{natbib}
\setcitestyle{super,sort&compress,comma}
\usepackage{svg}
\usepackage[version=3]{mhchem}
\usepackage{balance}
\usepackage{mathptmx}
\usepackage{sectsty}
\usepackage{graphicx}
\usepackage{lastpage}
\usepackage[format=plain,justification=justified,singlelinecheck=false,font={stretch=1.125,small,sf},labelfont=bf,labelsep=space]{caption}
\usepackage{float}
\usepackage{fancyhdr}
\usepackage{fnpos}
\usepackage[english]{babel}
\addto{\captionsenglish}{%
  
}
\usepackage{array}
\usepackage{droidsans}
\usepackage{charter}
\usepackage[T1]{fontenc}
\usepackage{setspace}
\usepackage[compact]{titlesec}
\usepackage{hyperref}
\usepackage{csvsimple}
\usepackage{longtable}
\usepackage{booktabs}
\usepackage{pgfsys}

\usepackage{chemformula} 
\usepackage[T1]{fontenc} 

\definecolor{cream}{RGB}{222,217,201}

\author{Vanessa Meschke}
\affiliation[Colorado School of Mines]
{Department of Physics, Golden, CO 80401}

\author{Prashun Gorai}
\affiliation[Colorado School of Mines]
{Department of Metallurgical and Materials Engineering, Golden, CO 80401}

\author{Vladan Stevanovi\'{c}}
\affiliation[Colorado School of Mines]
{Department of Metallurgical and Materials Engineering, Golden, CO 80401}

\author{Eric S. Toberer}
\affiliation[Colorado School of Mines]
{Department of Physics, Golden, CO 80401}

\email{etoberer@mines.edu}
\phone{303-273-3171}

\title[An \textsf{achemso} demo]
  {Search and Structural Featurization of Magnetically Frustrated Kagom{\'e} Lattices}

\abbreviations{IR,NMR,UV}

\begin{document}

\begin{tocentry}





\includegraphics{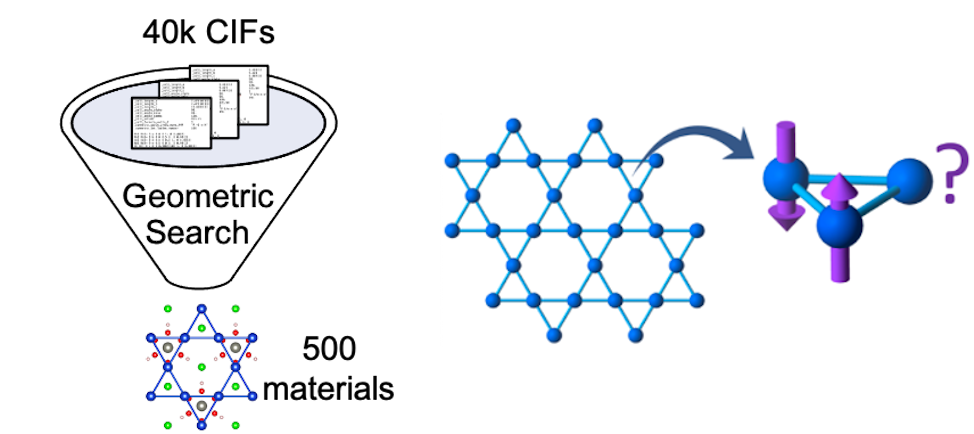}

\end{tocentry}

\begin{abstract}
  We have searched nearly 40,000 inorganic solids in the Inorganic Crystal Structural Database to identify compounds containing a transition metal or rare earth kagom{\'e} sublattice, a geometrically magnetically frustrated lattice, ultimately identifying $\sim$500 materials. A broad analysis of the chemical and structural trends of these materials shows three types of kagom{\'e} sheet stacking and several classes of magnetic complexity. Following the search and classification, we rapidly screen the magnetic properties of a subset of the materials using density functional theory to eliminate those that are unlikely to exhibit magnetic frustration. From the results of our computational screening, we rediscover six materials that have previously been explored for their low temperature magnetic behavior, albeit showing symmetry breaking distortions, spin glass behavior, or magnetic ordering. However, all are materials with antiferromagnetic behavior, which we correctly predict. Finally, we also report three materials that appear to be unexplored for their magnetic properties.
\end{abstract}
\clearpage

\section{Introduction}
Geometric magnetic frustration is often explored for the unique quantum phenomena it may create, ranging from enabling superconducting states\cite{SupercondMagFrust, Supercond2} to generating a quantum spin liquid (QSL)\cite{QSLColloquium}. While superconductivity needs no introduction, we'll briefly introduce QSLs, though more thorough reviews of QSLs are available for the reader\cite{QSL_States, QSLColloquium}. QSLs are a novel phase of matter that, despite the presence of strong magnetic interactions, do not exhibit magnetic ordering down to 0K\cite{Savary_Balents}. QSLs have yet to be proven to exist in real materials, but recent work has extensively focused on the characterization of materials with magnetically frustrated lattices\cite{Shores_HSM, QSLColloquium} and applying advancements in computational methods to better describe and predict the spin interactions of QSLs\cite{Xiang_4SM2}. After several decades of effort on both the experimental and theoretical fronts, few of the explored QSLs remain candidates. The lack of remaining QSL candidates arises from the relative ease of disproving a material to be a QSL as compared to the challenge of proving its existence, yielding a small and ever-narrowing search space for this exotic phase of matter. As such, the lack of candidate compounds limits both the ability to discover materials with exotic magnetic properties and learn from their aggregate behavior. The opportunity is thus for solid state chemistry to discover new materials that expand the structural landscape of possible QSLs. Herein, we ($i$) comprehensively identify materials with magnetic kagom{\'e} sublattices, ($ii$) review and explore their associated properties, and ($iii$) and identify candidates for further exploration.

While we choose to only explore the kagom{\'e} lattice in this work, there are numerous crystal lattices that exhibit geometric magnetic frustration. For example, the triangular lattice in 2D and the pyrochlore and hyperkagom{\'e} lattices in 3D all make arranging spins antiferromagnetically impossible. Additionally, other exchange-frustrated lattices, namely the square\cite{square_frustrated2, square_frustrated3, square_frustrated4} and honeycomb lattices\cite{honeycomb_frustrated, honeycomb_frustrated2, honeycomb_frustrated3, honeycomb_frustrated_exp, honeycomb_frustrated_exp_2} also exist. However, the square and honeycomb lattices' more restrictive requirements for J$_{i}$/J$_{j}$ ratios\cite{SquareFrustrated, j1j2_tighttolerance, honeycomb_frustrated3} and dependence of magnetic interactions up to third nearest neighbors\cite{honeycomb_frustrated4, 3nn_honeycomb} complicate a high-throughput search for frustration. As such, the kagom{\'e} lattice makes an excellent initial choice for a large scale search for new candidate QSLs due to its relative simplicity in spin coupling terms and its previous exploration in QSLs\cite{Kapella_Haydeeite, QSLColloquium}.

\begin{figure}[h]
\centering
  \includegraphics[height=3.8cm]{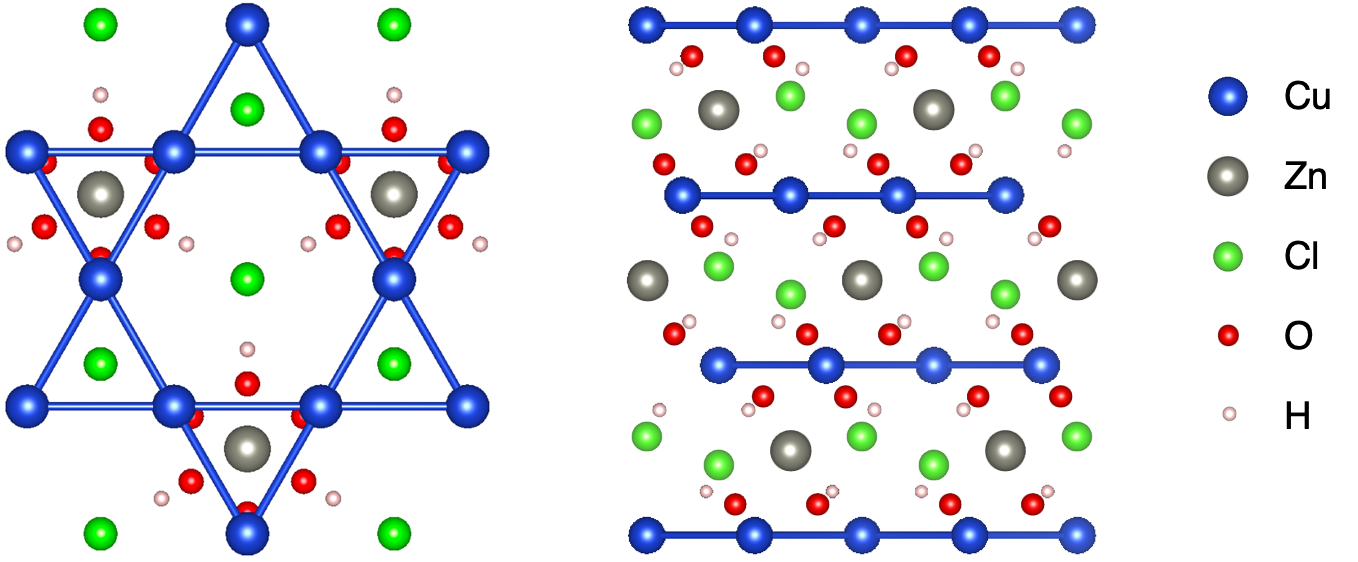}
  \caption{The crystal structure of herbertsmithite with the Cu$^{2+}$ kagom{\'e} sublattice highlighted in blue. On the left is one kagom{\'e} plane as seen when viewing the structure down the axis perpendicular to the planes. On the right is the staggering of the kagom{\'e} planes for a single unit cell of herbertsmithite.}
  \label{figure:HSM}
\end{figure}

In addition to its relative simplicity in coupling requirements, the kagom{\'e} lattice was selected for screening due to its relation to one of the strongest candidate QSLs to date: herbertsmithite (ZnCu$_{3}$(OH)$_{6}$Cl$_{2}$)\cite{Shores_HSM,braithwaite_mereiter_paar_clark_2004}. The crystal structure of herbertsmithite, with its Cu$^{2+}$ kagom{\'e} sublattice highlighted in blue, is shown in Figure \ref{figure:HSM}. Herbertsmithite has yet to be disproven for QSL behavior \cite{Han2012} with experimental measurements showing no magnetic ordering down to temperatures as low as 50mK.\cite{hsm_lowt1, hsm_lowt2}

Motivated by the ease of eliminating candidate QSLs and the challenge of generating new materials with frustrated lattices, we seek to define the structure space of existing materials with kagom{\'e} sublattices to accelerate the search for new QSLs. While Karigerasi, Wagner, and Shoemaker\cite{OtherKagomeList} have compiled a complete list of known 2D frustrated quantum magnets that can be searched, filtered, and output using a web-based frontend accessible from the Illinois Data Bank, the paper does not supply a categorization or overview of the magnetic or structural properties of the kagom{\'e}s. Additionally, other reviews of QSLs\cite{QSLColloquium,Savary_Balents} do an excellent job of outlining the challenges of this search and summarizing previously explored candidates, but they fail to provide a comprehensive assessment of all known kagom{\'e}s to date.

We begin with a structural search for compounds that are symmetry-ideal and nearly ideal kagom{\'e} sublattices from the Inorganic Crystal Structure Database (ICSD). Armed with this information, we explore the prototypes of materials with kagom{\'e} sublattices and propose a nomenclature for the describing these sublattices, focusing on structural trends for the magnetic species. After performing this classification, we generate a complete list of known kagom{\'e} prototypes regardless of their performance to date. Within a subset of this list, we rule out materials as candidate QSLs both by examining the literature for previously investigated kagom{\'e}s and analysis of spin-polarized density functional theory calculations. The final list of candidate materials can inspire more rigorous calculations and synthetic efforts as well as provide parent compounds for chemical mutations.

\section{Methods}
\begin{figure}[h]
  \centering
  \includegraphics[height=4cm]{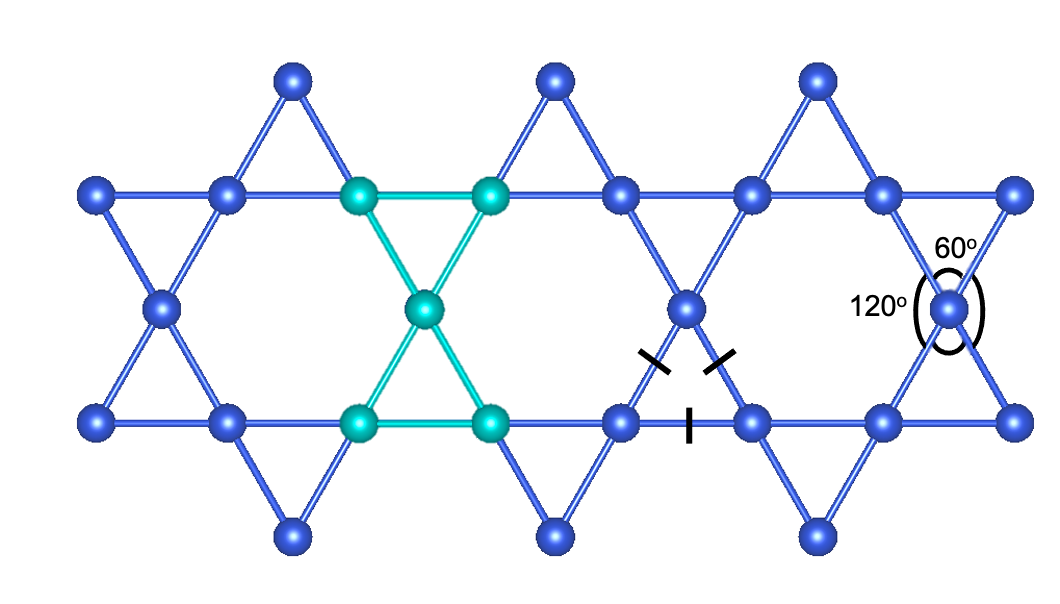}
  \caption{Geometric features of the kagom{\'e} lattice: four fold coordination highlighted in turquoise, equal side lengths of all triangles, and the two sets of 60$^{o}$ and 120$^{o}$ bond angles.}
  \label{figure:KagomeGeo}
\end{figure}

The initial dataset was collected by examining stoichiometric, ordered crystal structures documented in the ICSD for the geometric features of the kagom{\'e} lattice. In particular, the search screened for materials with 4-fold coordinated transition metal or rare earth atoms at unique Wyckoff positions. At each of these unique Wyckoff positions, materials were required to have equal bond distances between all neighbors and 2 each of 60, 120, and 180 degree bond angles (see Figure \ref{figure:KagomeGeo}). Additionally, all triangles lining the hexagon of the kagom{\'e} lattice needed to be equilateral. A tolerance of .3 \AA $\textrm{ }$was allowed on the bond distances and 5$^{o}$ for the bond angles. Finally, an additional search agnostic of Wyckoff position was also performed with the same geometric requirements as listed above. These geometric search constraints also included the hyperkagom{\'e} lattice in the search results.

The initial search yielded 497 unique compounds with magnetic atoms forming a kagom{\'e} sublattice at at least one Wyckoff position. Details of structural and chemical trends for these materials is detailed in the results section. A subset of 87 of the 497 compounds that contained no transition metal or rare earth atoms beyond the kagom{\'e} sublattice were selected for magnetic property screening using density functional theory (DFT). While DFT will not produce highly accurate magnetic properties for correlated materials, it is an efficient way to screen large datasets for non-magnetic compounds and poor candidate QSLs.

The magnetic moments of the calculation subset were calculated by performing spin-polarized calculations with GGA functionals and an automatic k-mesh using the Vienna Ab Initio Simulation Package (VASP). The high-throughput DFT calculations were performed using PyLada\cite{pylada}, a Python framework for the organizing and managing high-throughput first-principles calculations. Using Pylada, all spin configurations for supercells of the compounds were enumerated and their properties calculated.

To eliminate compounds with little potential as QSLs, the magnitude of the magnetic moments on the kagom{\'e} atoms of each of the compounds was first assessed. If a compound's kagom{\'e} sublattice showed no magnetic moment in any of its spin configurations, that compound was eliminated from further screening due to its predicted lack of magnetic properties. Following examination of the magnitude of the magnetic moments, the orientation of the moments in the lowest energy spin configuration for each compound was assessed. If a compound had only ferromagnetic (FM) interactions in its lowest energy spin configuration, it was eliminated from further assessment as it is unlikely a material that is predicted to be FM in DFT would be a candidate QSL. Finally, compounds with majority antiferromagnetic (AFM) interactions in their lowest energy spin configuration were most thoroughly examined for the energy differences between the lowest energy spin configuration and the FM configuration. While DFT is not the most accurate tool to compute the magnetic properties of a material, it is one of the only tools that provides a feasible means to rapidly screen even a fraction of the compounds in this dataset. More typical tools for predicting magnetic behavior, such as dynamic mean field theory, are simply too computationally expensive for a dataset of this size and for materials that may have more than 70 atoms in the unit cell.

The magnitude of the energy differences between spin configurations for structures with majority AFM interactions in their lowest energy spin configurations was investigated to approximate the strength of the coupling of the spins in the structure. While computational methods to determine the coupling coefficients for magnetic materials from DFT exist\cite{CrSiT3_Jterms, BiFeO3_Jterms}, they typically require using a spin Hamiltonian and accurately tuning the U parameter in the Hubbard model\cite{BiFeO3_Jterms} to produce accurate results. Additionally, the four-state method\cite{Xiang_4SM2, 4SM_Xiang, General_4SM_Sabani} also exists to compute the exchange coefficients, though it relies on knowing four precise magnetic configurations of the system to accurately map the energies of these configurations to the Hamiltonian for the system. As such, we instead examine the energy differences between the lowest energy AFM configuration and the FM configuration as a proxy to this coupling term to screen for compounds with interacting spins.

\section{Results and discussion}
\begin{figure}[h]
\centering
  \includegraphics[height=4.4cm]{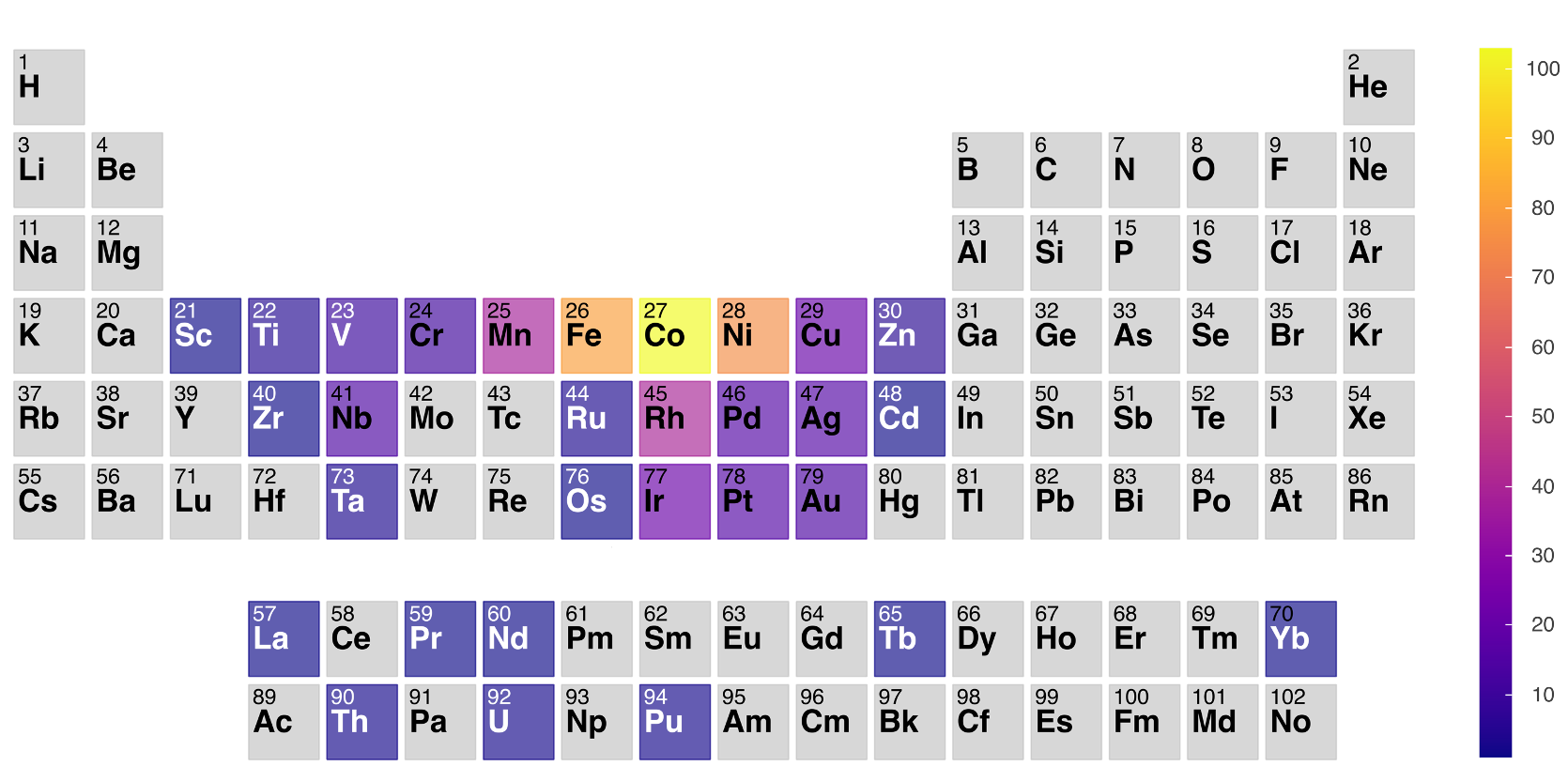}
  \caption{A heat map of the elements forming the kagom{\'e} sublattice in the 497 materials from the search. All transition metals and rare earths were considered in the search, so typically non-magnetic elements such as zinc and cadmium are shown in this heat map.}
  \label{figure:ptabletrend}
\end{figure}

To form the original dataset, $\sim$40,000 fully ordered, stoichiometric structures from the ICSD were screened for structural features of the kagom{\'e} lattice as detailed in the Methods. From this search, 497 compounds were found with a kagom{\'e} sublattice at a minimum of one Wyckoff position in the structure. A full list of these materials is detailed in a table in the Supplemental Information and the list is included as a CSV. Since the goal of this research is to build a comprehensive list of all known materials with a kagom{\'e} sublattice, the search included elements that are typically non-magnetic, such as zinc and cadmium. The chemical diversity of the elements forming the kagom{\'e} sublattice is highlighted as a heat map of the periodic table in Figure \ref{figure:ptabletrend}.  In terms of the counts of each element in the heatmap, the relatively weak showing of the more costly transition metals such as iridium and platinum is expected, especially considering their lack of representation in the ICSD. However, the comparatively high number of rhodium containing compounds is surprising, but appears to be due to three separate works\cite{Rh_1, Rh_2, Rh_3} that explore a total of 24 variants of rhodium kagom{\'e} sublattices. The large counts of iron, cobalt, and nickel are more expected given their prevalence in the ICSD.

Venturing beyond the elements forming the kagom{\'e} lattice, the compounds from our search can be categorized based on the elements surrounding the kagom{\'e} sublattice. Initially, the most apparent differentiation of these compounds are those that have additional transition metal or rare earth elements outside of the kagom{\'e} sublattice and those that do not, which we will designate as complex and simple kagom{\'e} compounds respectively. Complex kagom{\'e} compounds can be further divided into two subcategories. First, if the transition metal or rare earth atoms of the same species as the kagom{\'e} forming atom also appear outside of the kagom{\'e} sublattice, we will refer to these compounds as intrinsically complex kagom{\'e}s. Intrinsically complex kagom{\'e} compounds are unlikely to be QSLs due to the high likelihood of magnetic interactions between kagom{\'e} and non-kagom{\'e} atoms of the same species, breaking the kagom{\'e} sublattice's frustration. Alternatively, the transition metal or rare earth atoms outside the kagom{\'e} sublattice may be a different species than the kagom{\'e} forming atom. We refer to these materials as extrinsically complex kagom{\'e}s, and materials such as herbertsmithite, where zinc sits outside the Cu$^{2+}$ lattice, and YFe$_{6}$Ge$_{6}$, where yttrium atoms are outside the iron kagom{\'e} lattice, fall into this category. Extrinsically complex kagom{\'e} compounds may very well be candidate QSLs, especially if the atoms outside the kagom{\'e} sublattice carry no magnetic moment. A total 413 of the 497 compounds in the dataset are complex kagom{\'e}s, with 123 of those being intrinsically complex. The remainder are simple kagom{\'e}s.

Next examining the structural complexity of the 497 compounds in the data set, a plethora of crystal structure prototypes emerged. In total, 130 unique structure prototypes appeared, with 81 of those prototypes unique to a single compound in the dataset. Within the prototypes, MgFe$_{6}$Ge$_{6}$ was the most frequently appearing prototype, accounting for 13\% of the compounds' prototypes in this dataset. Beyond MgFe$_{6}$Ge$_{6}$, the CeCo$_{4}$B, Co$_{3}$GdB$_{2}$, Th$_{2}$Zn$_{17}$(filled), Ni$_{3}$Pb$_{2}$S$_{2}$, and ErIr$_{3}$B$_{2}$ prototypes also make strong appearances, in total comprising an additional 28\% of the appearing prototypes.

Within these prototypes, three alignments of the kagom{\'e} sheets were observed when viewing the kagom{\'e}s down the axis perpendicular to the planes. An example of each of the prototypes demonstrating the different alignments is shown in Figure \ref{fgr:staggering_ptypes}, and simplified examples of each are shown in the supplemental. First, the atoms of the kagom{\'e} sheets may align when looking down the direction perpendicular to the planes, which we refer to as aligned planes. Examples of prototypes with aligned planes would be the MgFe$_{6}$Ge$_{6}$ and Co$_{3}$GdB$_{2}$ prototypes. Aligned kagom{\'e} planes make up 310 of the 497 compounds in the dataset. If the kagom{\'e} planes are not aligned, the planes stack with lateral translation (shear) between the layers. The most common shear is a ($\frac{1}{3},\frac{1}{3}$) translation of the kagom{\'e} lattice vectors (e.g. the Ni$_{3}$Pb$_{2}$S$_{2}$ prototype). This shear keeps the overall six-fold rotational symmetry of the individual kagom{\'e} plane when extended into a stack. We refer to this stack as a symmetric shear, and it accounts for the majority of the sheared stacks observed in this data set and appears in 171 of the compounds. The final type of shear between planes is a ($\frac{1}{2}$,0) shear and is displayed by only the Cs$_{2}$Pt$_{3}$S$_{4}$ and K$_{2}$Pd$_{3}$S$_{4}$ prototypes. This shear does not preserve the six-fold axial symmetry and is referred to as asymmetric shear. Asymmetric shear is a fairly rare occurrence in this dataset, appearing in only 16 compounds. An example of each of the shears is shown in Figure \ref{fgr:staggering_ptypes} for a representative prototype of each category with its kagom{\'e} sublattice highlighted in blue, and a simplified graphic visualizing the three stacking categories is given in the supplemental. In addition to these alignment trends in kagom{\'e} plane stacking, all kagom{\'e} planes were found to either be equally spaced in a compound or to group in sets of two close planes with larger distances between the grouped planes. The plane alignment and spacing is included as data in a CSV.
\begin{figure}[h]
\centering
  \includegraphics[height=4.5cm]{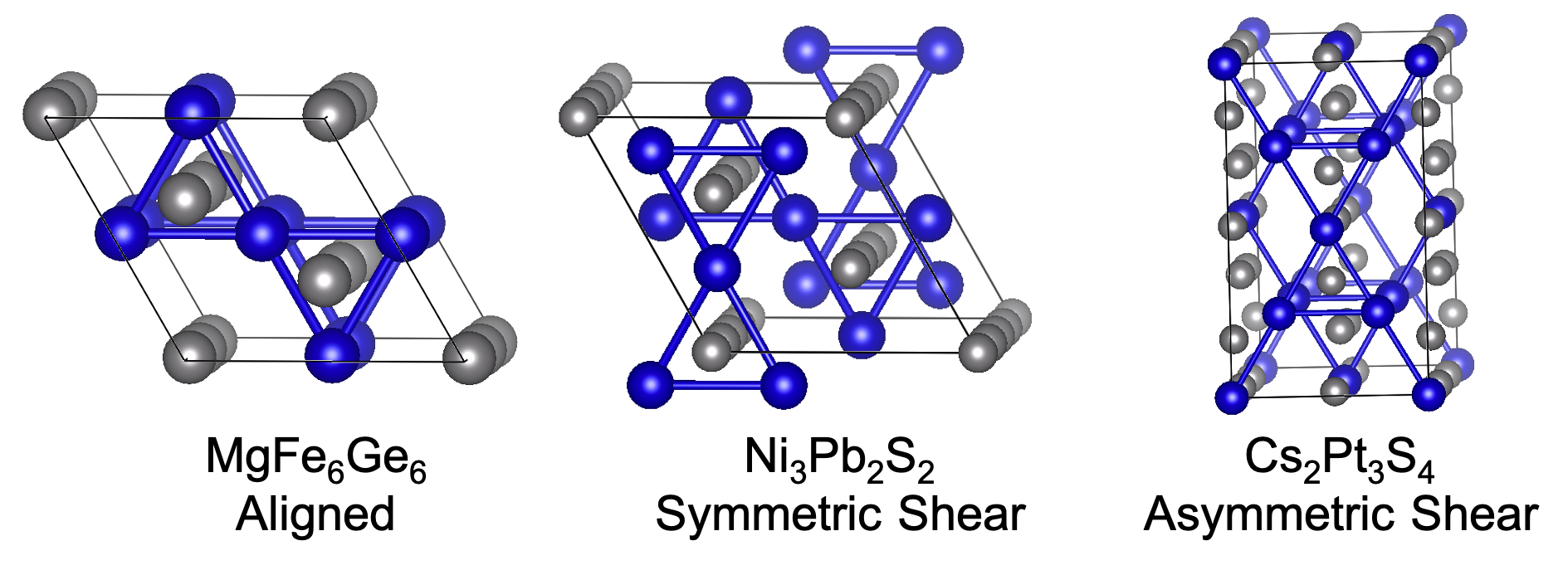}
  \caption{Representative prototype for each type of kagom{\'e} plane alignment. From left to right, the kagom{\'e} planes are aligned in MgFe$_{6}$Ge$_{6}$, symmetrically sheared for Ni$_{3}$Pb$_{2}$S$_{2}$, and asymmetrically sheared for Cs$_{2}$Pt$_{3}$S$_{4}$. In each prototype, the kagom{\'e} sublattice is highlighted in blue.}
  \label{fgr:staggering_ptypes}
\end{figure}

Further examining the structural trends in this our dataset, we next examine the specific Wyckoff positions that form the kagom{\'e} sublattice. Examples of each of the following Wyckoff position dependence for kagom{\'e} sublattice formation are given in the supplemental. In this data set, we found that kagom{\'e}s are either formed by a single Wyckoff position or multiple Wyckoff positions. 392 compounds were formed from only one Wyckoff position for all planes. For the remaining compounds whose kagom{\'e}s planes are composed of multiple Wyckoff positions, two the different Wyckoff positions either resided in different planes or the same plane. For example, if  multiple Wyckoff positions correspond to  separate planes of the kagom{\'e} sublattice, one Wyckoff position may form a kagom{\'e} layer at z = 0, with another Wyckoff position forms the kagom{\'e} layer at z = 0.5 for kagom{\'e} planes that stack in the z direction. This is the case for the Ce$_{3}$Co$_{11}$B$_{4}$ prototype. Alternatively, multiple Wyckoff positions may exist in a single kagom{\'e} plane, and all planes are consist of multiple Wyckoff positions. For example, materials in the ErIr$_{3}$B$_{2}$ prototype exhibit this type of combined-Wyckoff kagom{\'e} plane. For compounds whose kagom{\'e} planes are formed from multiple Wyckoff positions, 56 had different Wyckoff positions at different layers, whereas 49 compounds had kagom{\'e} where each plane consisted of multiple Wyckoff positions.

Moving beyond the structural and chemical trends of the full data set, the magnetic properties of a subset of compounds were also investigated to begin building insights to the connections between structure, chemistry, and magnetic behavior. A subset of 87 compounds containing no transition metals or rare earths beyond the kagom{\'e}-forming species were selected for an initial screening with DFT and a literature review to validate predictions of magnetic behavior. The compounds in this set spanned a diverse set of prototypes and transition metal or rare earth atoms forming the kagom{\'e} sublattice, each of  which are featured Figure \ref{fgr:strc72_ptypes}, and a full list of these compounds is included as a table in the supplemental and as a CSV. For Figure \ref{fgr:strc72_ptypes}, a bin for each prototype is listed along the y axis of the heat map, and the options for transition metal or rare earth atoms that form the kagom{\'e} sublattice is on the x axis. Each box of the heat map details how many times a given prototype appeared in  the calculation set with the corresponding transition metal or rare earth atom forming its kagom{\'e} sublattice. For example, the CoSn prototype appeared three times in the calculation set, which is noted on the histogram to the right of the heat map. For the three times the CoSn prototype appeared, it had 3 different transition metals forming its kagom{\'e} sublattice: iron, cobalt, and rhodium. The histogram to the right details how many times each prototype appeared in the calculation set. Considering the structural trends, most protoypes are unique to a specific compound in this dataset, and most kagom{\'e} sublattices are formed from Fe, Ni, Cu, and Au. For one of the most well represented prototype, Ni$_{3}$Pb$_{2}$S$_{2}$, a historical bias in the data is present due to the thorough exploration of the shandites' half-metallic ferromagnetic behavior\cite{Weihrich_HalfAPV, Weihrich_XtalElectricShandites, Ni3Sn2S2_nonmag, Kassem_ShanditeAlloying}.
\begin{figure}[h]
\centering
  \includegraphics[height=5.5cm]{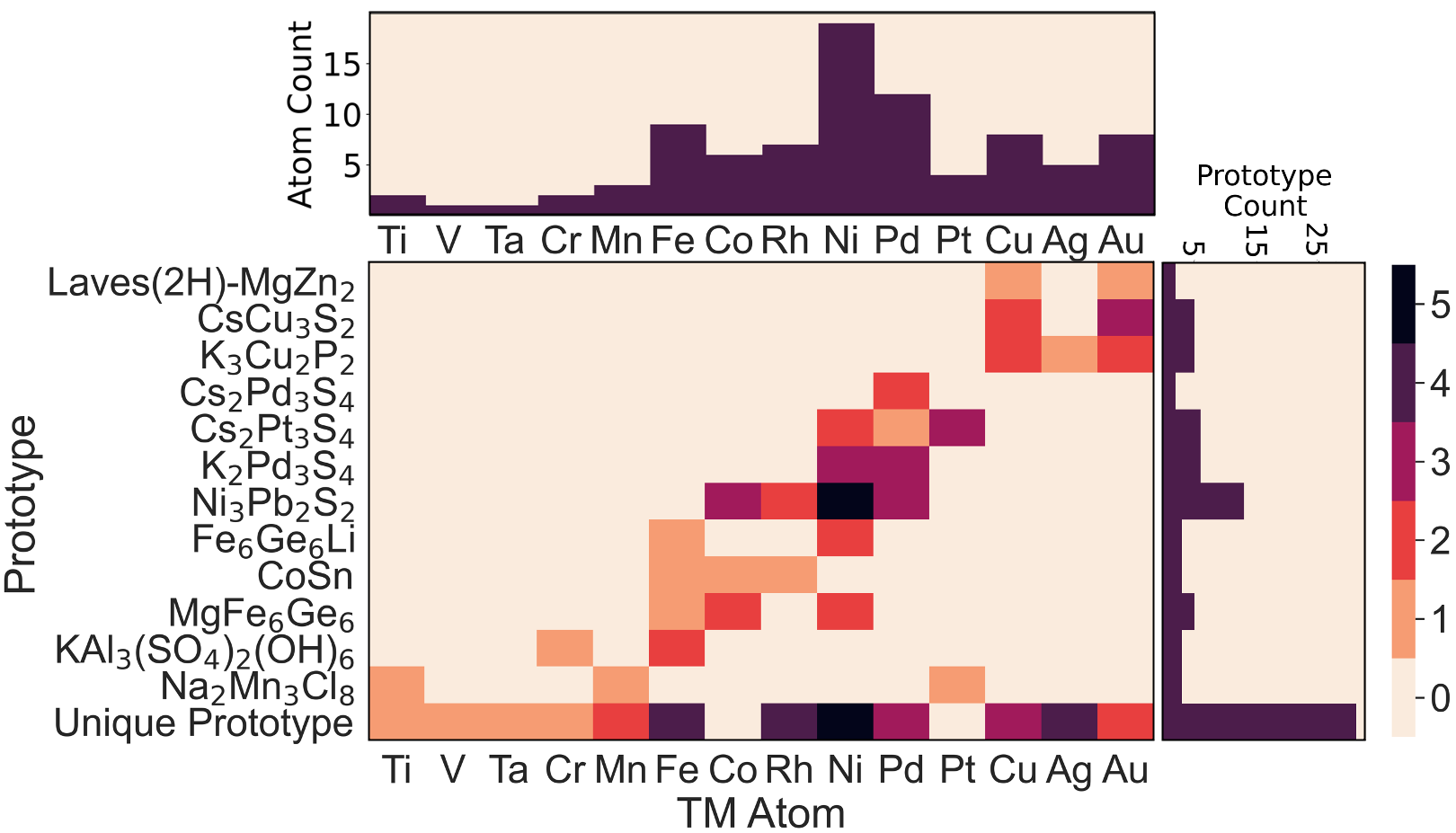}
  \caption{A heat map of which prototype and transition metal or rare earth atom combinations appeared for the compounds investigated with DFT. Each box of the heat map details how many times a given prototype appeared in the calculation set with the corresponding transition metal or rare earth atom forming its kagom{\'e} sublattice. }
  \label{fgr:strc72_ptypes}
\end{figure}

\begin{figure}[h]
\centering
  \includegraphics[height=6.5cm]{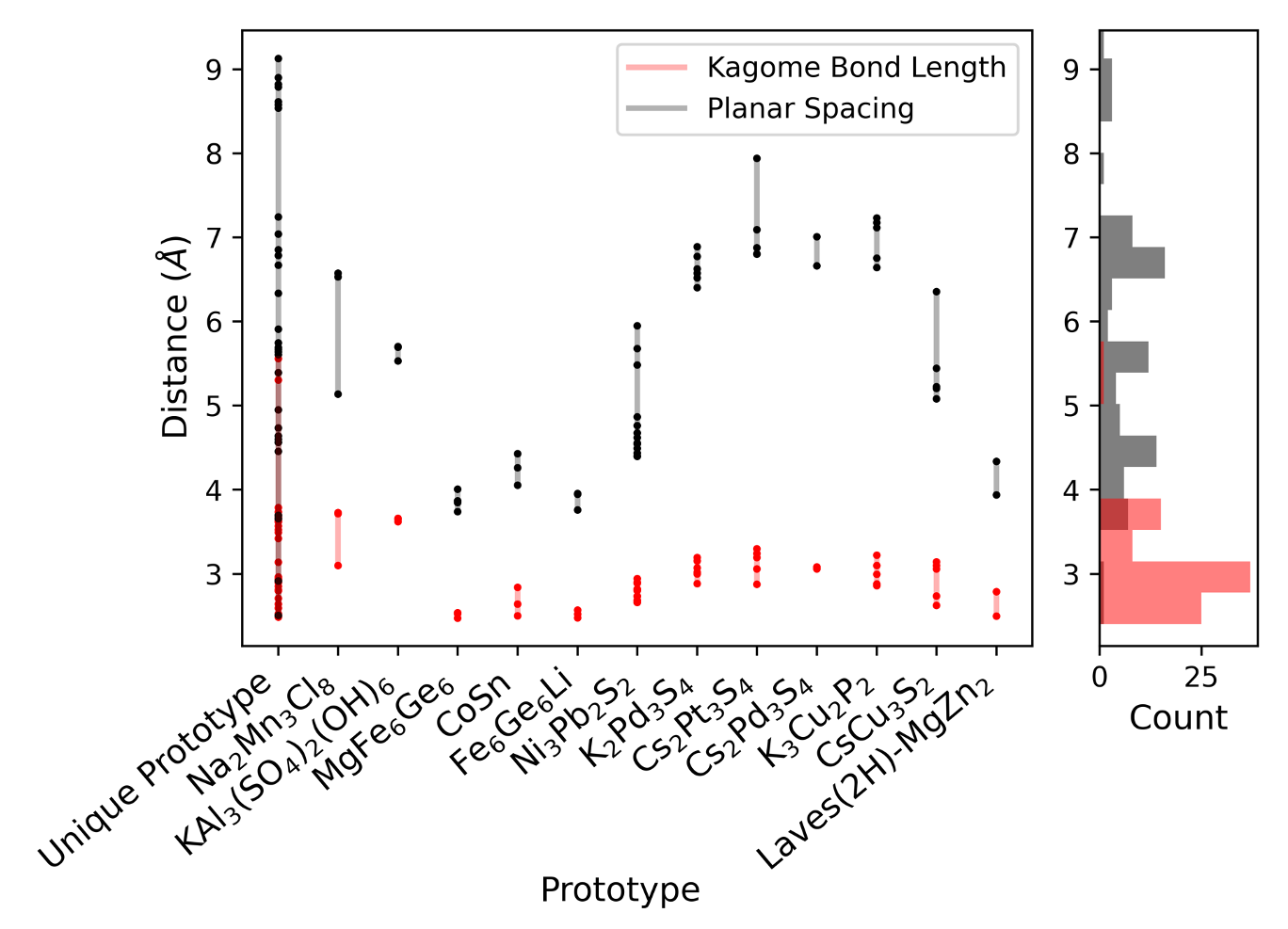}
  \caption{Comparison of the kagom{\'e} bond distance (red) and planar spacing (black) for the calculation set. The values of the bond distances and planar spacings are binned in the histogram to the right of the scatter plot.}
  \label{fgr:bond_planes}
\end{figure}

Going beyond the trends in kagom{\'e} atom type, Figure \ref{fgr:bond_planes} details the comparisons of the distances between nearest neighbor kagom{\'e} atoms, which will be referred to as the kagom{\'e} bond distance, and the distances between the planes of the kagom{\'e}s for each prototype. A separate version of this plot for all compounds in the dataset is included in the supplemental.  The range of both kagom{\'e} bond distance and planar spacing of the `Unique Prototype' bin particularly highlights the diversity in structural trends for compounds with a kagom{\'e} sublattice and shows promise for the discovery of new kagom{\'e} sublattices given their ability to exist around many additional types of atoms between their layers. For the entire dataset, the mean kagom{\'e}-kagom{\'e} bond length is 3.05\AA ($\sigma$: 0.52 \AA) and the average distance between kagom{\'e} planes is 5.69 \AA ($\sigma$: 1.48 \AA). However, the `Unique Prototype' bin displays a much large deviation in both bond lengths and planar spacings. If the unique prototype bin is excluded, more consistency in the kagom{\'e} bond distance is found (avg: 2.92\AA$\textrm{ }$, $\sigma$: 0.32\AA), and the distance separating the kagom{\'e} planes varies between prototypes and within a given prototype (avg: 5.47\AA, $\sigma$: 1.19\AA).

Having established structural and chemical trends in the compounds found, we move forward with computational screening and a literature review of the 87 compounds in the calculation set. To investigate the quality of these materials as candidate QSLs, the magnitude and orientation of the magnetic moments and the energy of the various spin configurations were collected from the DFT calculations. For the 87 compounds studied, 62 were found to be non-magnetic with DFT. These 62 compounds were mostly composed of the shandites in the Ni$_{3}$Pb$_{2}$S$_{2}$ prototype group as well as many of the uniquely appearing prototypes. The remaining 25 compounds had magnetic moments greater than 0.5 $\mu_{B}$ per atom in the structure and are classified more thoroughly below.

To assess the overall validity of these magnetic moment calculations, calculation results were compared with experimental findings in the literature. Overall, our calculations find good agreement with experiment. For example, the shandite Co$_{3}$Sn$_{2}$S$_{2}$ has been experimentally found to be ferromagnetic below 175K,\cite{Co2Sn2S2_fm} and its Ni analog, Ni$_{3}$Sn$_{2}$S$_{2}$, is non-magnetic,\cite{Ni3Sn2S2_nonmag} both of which are confirmed by our calculations. Additionally, KV$_{3}$Ge$_{2}$O$_{9}$ is reported AFM in literature,\cite{KV3Ge2O9_AFM} a result that is also confirmed by our calculations, though with a small energy differences between spin configurations in our calculations.

For the 25 materials with sizable magnetic moments, 10 had a ferromagnetic (FM) in plane and between plane spin configuration as their lowest energy. Of the remaining 15 compounds, 3 compounds had FM interactions in-plane with antiferromagnetic (AFM) interactions between planes. The final 12 compounds had AFM interactions as their lowest energy configurations, and of these 12 compounds, five had small (< 5meV/kagom{\'e} forming atom) differences in energy between the lowest energy configuration with mostly AFM interactions and the FM configuration. The remaining 7 compounds with large energy differences between spin configurations and ground state that shows mostly AFM interactions are explored more thoroughly below. Additionally, we highlight two compounds (Na$_{2}$Mn$_{3}$Cl$_{8}$ and Cu$_{3}$Pb(AsO$_{4}$)$_{2}$(OH)$_{2}$) whose energy differences are smaller than this cut off but appear to have no literature regarding their magnetic properties.

\subsection{Jarosites: \textit{A}Fe$_{3}$(OH)$_{6}$(SO$_{4}$)$_{2}$$\textrm{ }$ (\textit{A} = K, Na, H$_{3}$O)}

Our dataset contained four compounds with the KAl$_{3}$(SO$_{4}$)$_{2}$(OH)$_{6}$ prototype: KFe$_{3}$(OH)$_{6}$(SO$_{4}$)$_{2}$, NaFe$_{3}$(OH)$_{6}$(SO$_{4}$)$_{2}$, (H$_{3}$O)Fe$_{3}$(OH)$_{6}$(SO$_{4}$)$_{2}$, and KCr$_{3}$(SO$_{4}$)$_{2}$(OH)$_{6}$. KCr$_{3}$(OH)$_{6}$(SO$_{4}$)$_{2}$ is a mineral known as alunite, which we predicted to have mostly AFM interactions in its lowest energy configuration. While  KCr$_{3}$(OH)$_{6}$(SO$_{4}$)$_{2}$ showed relatively small energy differences between its lowest energy and its FM configuration (4 meV/Cr atom), the remaining three members of this prototype showed more promise as candidate QSLs. KFe$_{3}$(OH)$_{6}$(SO$_{4}$)$_{2}$, NaFe$_{3}$(OH)$_{6}$(SO$_{4}$)$_{2}$, and (H$_{3}$O)Fe$_{3}$(OH)$_{6}$(SO$_{4}$)$_{2}$) are compounds belonging to the mineral group of jarosites, and we will focus the rest of this section on this class of materials.

Jarosites are a naturally occurring hydrous sulfate mineral with a composition following the formula AFe$_{3}$(OH)$_{6}$(SO$_{4}$)$_{2}$, (A = Na, K, Rb, H$_{3}$O, Pb, or other metals or molecules), where the Fe atoms form the kagom{\'e} sublattice in the crystal \cite{Basciano2007}. Previous investigations into both naturally occurring and synthetic jarosites has shown they typically exist with large concentrations of vacancies on the kagom{\'e} sublattice\cite{Grohol2003,Basciano2007}. Despite this large fraction of vacancies, jarosites are considered to be nearly ideal Heisenberg antiferromagnets, especially when the vacancy concentration is minimized\cite{Harrison2004}.

The original jarosite for which this group of minerals derives its name has the formula KFe$_{3}$(OH)$_{6}$(SO$_{4}$)$_{2}$, where the Fe atoms form the kagom{\'e} sublattice. From DFT, we find jaroiste's lowest energy Ising spin configuration, which mostly consists of AFM interactions, and the FM configuration to have an energy difference of 90 meV/Fe atom with a magnetic moment of 4.2 $\mu_{B}$ per Fe atom. Next we examine a variation on jarosite where the K atoms surrounding the kagom{\'e} lattice are swapped for Na, creating natrojarosite (NaFe$_{3}$(OH)$_{6}$(SO$_{4}$)$_{2}$). We predict natrojarosite to have a magnetic moment of 4.2 $\mu_{B}$ per Fe atom. Additionally, the lowest energy configuration with mostly AFM interactions and FM spin configurations varied by 132 meV/Fe atom for natrojarosite. Finally, hydronium jarosite ((H$_{3}$O)Fe$_{3}$(OH)$_{6}$(SO$_{4}$)$_{2}$) showed similar results to its previous mineral family members, showing a magnetic moment of 4.2 $\mu_{B}$ per Fe atom and an energy difference of 105 meV per Fe atom from the mostly-AFM to the FM spin configurations.

Most jarosites have been experimentally shown to undergo long range magnetic ordering between 50-65K. In particular, KFe$_{3}$(OH)$_{6}$(SO$_{4}$)$_{2}$ orders at 65 K\cite{jarosite_inami} and NaFe$_{3}$(OH)$_{6}$(SO$_{4}$)$_{2}$ at 50 K\cite{Wills2000}. Hydronium jarosite, however, displays a break from the ordering trend and has been shown to be spin glass with a glass transition temperature of 15K\cite{hydronium_jarosite}. These experimental results eliminate all of the jarosites for QSL behavior. However, future investigations into the jarosites could model Wills'\cite{Wills2000,DopingJarosites} approach to taking advantage of the large vacancy concentration on the Fe sites and investigate the magnetic properties of doped jarosites to further illuminate their magnetic behavior .

\subsection{Cs$_{2}$KMn$_{3}$F$_{12}$}
From DFT, Cs$_{2}$KMn$_{3}$F$_{12}$ shows promise as a QSL candidate with a magnetic moment of 3.7 $\mu_{B}$/Mn atom and a difference of 200 meV/Mn atom between the mostly-AFM and FM spin configurations. This compound and those similar in their chemistry and structure have been previously investigated for its magnetic properties. While compounds such as Cs$_{2}$LiMn$_{3}$F$_{12}$ and Cs$_{2}$NaMn$_{3}$F$_{12}$ order at 2.1 K and 2.5 K, respectively, the ordering temperature of Cs$_{2}$KMn$_{3}$F$_{12}$ appears to order around 7 K\cite{Cs2Cu3F12Sn1_AFMOrder, Cs2F12K1Mn3_OrigMag}. However, it is unknown if the ordering in this sample is long-range or short range and the exact nature of the ordering should be investigated experimentally.

\subsection{Cs$_{2}$SnCu$_{3}$F$_{12}$}
Similar in composition to Cs$_{2}$KMn$_{3}$F$_{12}$ but quite different in structure, our predictions initially show Cs$_{2}$SnCu$_{3}$F$_{12}$ to be a candidate for QSL behavior. With a magnetic moment of 0.77 $\mu_{B}$/Cu atom and an energy difference of 12 meV per Cu atom between the mostly AFM and the FM states, Cs$_{2}$SnCu$_{3}$F$_{12}$ appears promising  from our predictions. However, literature shows that Cs$_{2}$SnCu$_{3}$F$_{12}$ experiences long range magnetic ordering at 17K and structural distortions are reported around 185 K\cite{Cs2SnCu3F12_longrangeorder,Cs2SnCu3F12_longrangeorder2}, experimentally eliminating it as a candidate QSL.

\subsection{Na$_{2}$Ti$_{3}$Cl$_{8}$ and Na$_{2}$Mn$_{3}$Cl$_{8}$}
In calculation, Na$_{2}$Ti$_{3}$Cl$_{8}$ showed promise for a QSL with a magnetic moment of 1.7 $\mu_{B}$ per Ti atom in the structure and an energy difference of 390 meV/Ti atom in the structure from the lowest energy, mostly-AFM configuration to the FM configuration. Despite its computational promise, this material has been previously investigated for its magnetic properties and has been shown to undergo a Peierl's-type distortion at 200K, forming Ti trimers,\cite{NaTiCl_distort} experimentally eliminating it as a candidate QSL. Calculations were also performed on the Mn analog of this compound, Na$_{2}$Mn$_{3}$Cl$_{8}$, which was found to have a magnetic moment of 4.5 $\mu_{B}$ per Mn atom. However, Na$_{2}$Mn$_{3}$Cl$_{8}$ shows very small energy differences (1.6 meV/Mn atom) between the mostly-AFM and FM spin configurations, with the AFM being lower. Magnetic measurements of Na$_{2}$Mn$_{3}$Cl$_{8}$ have been made previously, finding no distortions or magnetic ordering down to 100 K\cite{Na2Mn3Cl8Mag}. However, this compound should be further investigated experimentally to determine if the distortions observed in Na$_{2}$Ti$_{3}$Cl$_{8}$ persist with swaps of the kagom{\'e}-forming atom and to determine an ordering temperature if one exists.

\subsection{Corkite, PbFe$_{3}$(SO$_{4}$)(PO$_{4}$)(OH)$_{6}$}
Similar in structure to the jarosites, we also find the compound corkite (PbFe$_{3}$(SO$_{4}$)(PO$_{4}$)(OH)$_{6}$) to have a spin configuration with mostly AFM interactions as its lowest energy state in calculation. With a magnetic moment of 4.2 $\mu_{B}$ per Fe atom and an energy difference of 80 meV per Fe atom between the lowest energy, mostly-AFM and FM spin configurations, corkite shows a remarkable similarity to the jarosites in DFT.

Corkite is another naturally occurring mineral that can be found in acidic mine runoff\cite{Corkite_wastedumps}. While corkite's structure has been extensively studied for determining the location of its phosphate and sulfate groups and to cope with the challenges of creating a high quality crystalline sample\cite{Corkite_wastedumps, Corkite_struct, GIUSEPPETTI1987}, little about its properties are reported in the literature other than its light brown color and its potential application in identifying economically valuable ores for mining\cite{Corkite_indicator}. Magnetic measurements of corkite should be made to determine if this mineral orders as the jarosites do.

\subsection{Bayldonite, Cu$_{3}$Pb(AsO$_{4}$)$_{2}$(OH)$_{2}$}
Though it has a smaller energy difference between its spin configurations than previously highlighted compounds in this section, we choose to highlight Cu$_{3}$Pb(AsO$_{4}$)$_{2}$(OH)$_{2}$ because its lack of magnetic data in the literature. Cu$_{3}$Pb(AsO$_{4}$)$_{2}$(OH)$_{2}$, also known by the mineral name bayldonite, was shown to have a configuration dominated by AFM interactions as its lowest energy configuration with a magnetic moment of roughly 0.7 $\mu_{B}$/Cu atom and large energy differences between the lowest energy and the FM spin configurations (on the order of 4.2 meV/Cu atom). To the best of our knowledge, this compound has not been investigated for its magnetic properties and should be explored experimentally to learn more about its magnetic behavior.

\begin{figure}[h]
\centering
  \includegraphics[height=5cm]{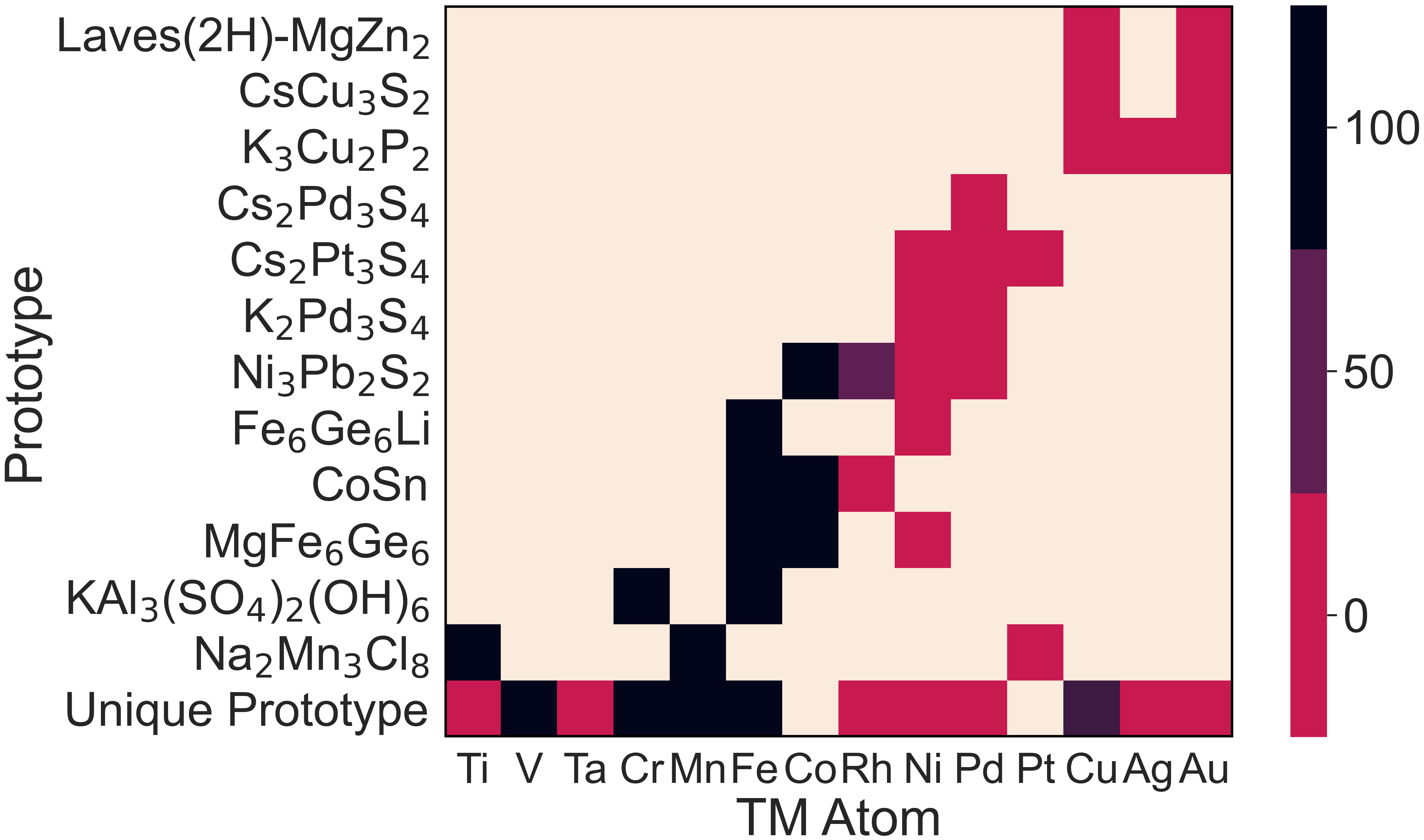}
  \caption{Percent of elements with a magnetic moment for each prototype for the calculation subset of kagom{\'e}s. Ni, Pd, Pt, Ag, and Au never produced moments regardless of prototypes, while Cr, Mn, Fe, and Co always did.}
  \label{fgr:tm_atom_presence}
\end{figure}

In the subset of materials DFT was performed on, we analyzed the proclivity for the transition metal or rare earth elements produce a magnetic moment as a function of the prototype each appeared in. Our findings are summarized in Figure \ref{fgr:tm_atom_presence} by giving a percentage of each transition metal or rare earth in each prototype that had a magnetic moment from DFT.
As expected, certain transition metals and rare earths are more prone to displaying a magnetic moment from calculation. Cr, Mn, Fe, and Co were always predicted to have a magnetic moment, while Ni, Pd, Pt, Ag, and Au never did. On the more ambiguous end of the spectrum are Ti, Cu, and Rh, which only produced moments when appearing in specific prototypes, each of which were expected given the charge counting. Given the small number of compounds investigated in each prototype, determining which prototypes are most likely to produce a magnetic properties from this dataset is unwise. Regardless, each of these prototypes can be manipulated to expand the possible structure space available for candidate QSLs. By swapping the various transition metal or rare earths forming the kagom{\'e} lattice, trends between structure, chemistry, and magnetism can be further elucidated while providing more options to search for candidate QSLs.

\section{Conclusion}
Despite hundreds of kagom{\'e} sublattice-containing compounds in known materials, there was a dearth of knowledge on their identification as well as structural and magnetic properties. By creating this dataset, we have been able to show that the kagom{\'e} sublattices appear in a chemically and structurally diverse set of materials. We report kagom{\'e} sublattices formed from 30 different transition metal or rare earths and 130 unique prototypes with variety of spacings between kagom{\'e} atoms both in plane and out of plane, as well as three different trends for stacking of the kagom{\'e} planes.

Additionally, we demonstrate of a method of computationally screening materials for their magnetic properties, particularly magnetic frustration, which is necessary for enhancing the search for candidate QSLs. As is elucidated by this search, there is a wide chemical and structural space for kagom{\'e} sublattices to exist in and it is nowhere near fully explored. Finally, we predict nine candidate materials for possible QSL behavior from the results of our calculations. Of these nine materials, six (the jarosites, Na$_{2}$Ti$_{3}$Cl$_{8}$, Cs$_{2}$KMn$_{3}$F$_{12}$, and Cs$_{2}$SnCu$_{3}$F$_{12}$) have been experimentally eliminated as candidate QSLs, leaving Na$_{2}$Mn$_{3}$Cl$_{8}$, corkite, and bayldonite as promising candidate QSLs to be investigated experimentally following this work.

\begin{acknowledgement}

V. Meschke acknowledges this material is based upon work supported by the National Science Foundation Graduate Research Fellowship Program under Grant No. 1646713. EST acknowledges NSF award 1555340. The research was performed using computational resources sponsored by the Department of Energy's Office of Energy Efficiency and Renewable Energy and located at the National Renewable Energy Laboratory.

\end{acknowledgement}

\section*{Conflicts of Interest}
There are no conflicts to declare.

\begin{suppinfo}

The following files are available free of charge as supporting information:
\begin{itemize}
  \item calculation\_set.xls: Spreadsheet listing the compounds DFT was used to screen magnetic properties on, as well as some key results of the calculations.
  \item full\_kagome\_dataset.xls: Spreadsheet listing all found kagom{\'e} compounds from the search of the ICSD.
\end{itemize}

\end{suppinfo}

\bibliography{rsc.bib}

\end{document}